\documentstyle[aps,prl,twocolumn,psfig]{revtex}
\begin{document}
\title{The Potential Energy Landscape and Mechanisms of Diffusion in Liquids}
\author{T. Keyes and J. Chowdhary}
\address{{\it Department of Chemistry, Boston University, Boston, MA 02215}}
\date{(\today)}
\maketitle
\begin{abstract}
The mechanism of diffusion in supercooled liquids is investigated from the potential energy {\it landscape} point of view, with emphasis on the crossover from high- to low-$T$ dynamics over the range $T_{A} \geq T \geq T_{c}$. Molecular dynamics simulations with a time dependent mapping to the associated local mininum or {\it inherent structure} (IS) are performed on unit-density Lennard-Jones (LJ). New dynamical quantities introduced include $r2_{is}(t)$, the mean-square displacement (MSD) within a basin of attraction of an IS, $R2(t)$, the MSD of the IS itself, and $g(t)$, the distribution of IS waiting times. The configuration space is treated as a composite of the contributions of cooperative local regions, and a method is given to obtain the physically meaningful $g_{loc}(t)$ and mean waiting time $\tau_{loc}$ from $g(t)$. A new understanding of the crossover is obtained in terms of $r2_{is}(t)$ and $\tau_{loc}$. At intermediate $T$, $r2_{is}(t)$ posesses an interval of linear $t$-dependence allowing calculation of an intrabasin diffusion constant $D_{is}$. Near $T_{c}$, where intrabasin diffusion is well established for $t < \tau_{loc}$, diffusion is {\it intrabasin dominated} with $D = D_{is}$; {\it $D$ may be calculated within a basin}. Below $T_{c}$, $\tau_{loc}$ exceeds the time, $\tau_{pl}$, needed for the system to explore the basin, indicating the action of barriers at the border; $\tau_{loc} = \tau_{pl}$ is a criterion for a transition to activated hopping. Intrabasin diffusion provides a means of confinement not involving barriers and plays a key role in dynamics above $T_{c}$. The distinction between motion among the IS (IS dynamics) below $T_{c}$ and {\it saddle, or border dynamics} above $T_{c}$, where the system is always close to one of the saddle-barriers connecting the basins and IS boundaries are closely spaced and easily crossed, is discussed. A {\it border index} is introduced based upon the relation of $R2(t)$ to the conventional MSD, and shown to vanish at $T \sim T_{c}$. It is proposed that intrabasin diffusion is a manifestation of saddle dynamics.
\end{abstract}

\section{Introduction; Diffusion and the Potential Energy Landscape \label{sec-intro}}
Knowledge of the physical mechanism of diffusion is essential to an understanding of the dynamical complexities of supercooled liquids. A mechanism should include the collective motions composing the elementary diffusive step, any correlations among the steps, and a statement about whether barrier crossing is important. If so, distributions of barrier heights, 'reaction coordinates' for barrier crossing, and details of the relevant saddle-barriers are significant. Valuable information is encoded in the $T$-dependence of the self diffusion coefficient $D$ and of the viscosity $\eta$ (the Stokes-Einstein law, $D \sim T/\eta$, holds until very close to the glass transition at $T_{g}$). For most normal liquids $D$ exhibits Arrhenius $T$-dependence. In \cite{angell0} strong liquids this continues all the way down to $T_{g}$, while super-Arrhenius, with the apparent activation energy $E_{A}(T)$ increasing with decreasing $T$, sets in before $T_{g}$ in \cite{angell0} fragile liquids. Some exceptions classified as 'weak' by Kivelson et. al \cite{dk}, including simple atomic models such as Lennard-Jones (LJ), show a normal or upper-supercooled range of power-law $D(T)$. Clearly the mechanism of diffusion is $T$-dependent, and a straightforward conclusion is that activated potential-energy barrier crossing, or {\it hopping}, plays an important role. In contrast to chemical reactions where the Arrhenius law was first introduced, however, the barriers and reaction coordinates for diffusion have proved elusive. To further complicate matters, free energy or 'entropic' barriers, associated with the difficulty of finding a low-energy pathway, may also be important. It is not straightforward to infer a mechanism from $D(T)$, and thus there has been considerable speculation.

Adam and Gibbs suggested \cite{ag} that liquids may be decomposed into roughly independent {\it cooperatively rearranging regions} (CRR) containing $z^{*}$ particles, with super-Arrhenius arising from the growth of the CRR, and thus the activation free energy, with decreasing $T$. An influential paper by Goldstein \cite{gold} formulated dynamics in terms of the $3N$-dimensional potential energy $U$-surface, or {\it landscape}. He proposed the existence of distinct high and low-$T$ dynamical regimes, separated by a crossover temperature. At low $T$ the system stays for times long compared to a vibrational period in the basins associated with the local minima of the landscape, and diffusion is governed by infrequent hopping of the saddle-barriers connecting adjoining basins. Perhaps the rearrangements of the CRR correspond to reaction coordinates on the landscape. At high-$T$ a different, freer motion is presumed to occur. The onset of the low-$T$ mechanism is sometimes described as a 'transition to hopping', but if the high-$T$ diffusion constant is Arrhenius, implying activation, the change must be more subtle. The idea of a crossover is central to current thinking about supercooled liquids. We will pursue the landscape point of view in this paper, so the mechanism of diffusion is naturally discused in terms of the topology. 

Goldstein's picture has been extended to involve two characteristic temperatures, meaning that the mechanism changes gradually over a crossover range $T_{A} \geq T \geq T_{c}$. Interesting features of supercooled dynamics such as super-Arrhenius and stretched exponential decay of time correlations first appear at $T_{A}$ \cite{angell} and $D(T)$ extrapolates to zero from above, via power-law fit, at the mode-coupling \cite{goetze} temperature $T_{c}<T_{A}$.  Note that $T_{c}$, defined by the extrapolation, seems a useful quantity regardless of the validity of mode-coupling theory.  An alternative formulation, the \cite{dk1} 'frustration limited domain'  theory of Kivelson and co-workers, introduces an upper crossover temperature $T^*$ at which collective dynamics first becomes important. Still lower \cite{angell0,dk} lie $T_{g}$, the Kauzmann temperature $T_{K}$ and the VTF $T_{0}$, but this paper is concerned with temperatures from $T_{A}$ to somewhat under  $T_{c}$. It is difficult to obtain well-equilibrated computer simulation data below $T_{c}$, and impossible to approach $T_{g}$.

Stillinger and Weber showed \cite{sw,fssci} how to obtain an explicit description of motion over the landscape in a molecular dynamics simulation, by mapping a $3N$-dimensional liquid configuration ${\bf r}(t)$ to the associated minimum ${\bf R}(t)$, or {\it inherent structure} (IS), with frequent steepest-descent minimizations (faster conjugate-gradient minimizations are equivalent \cite{tkjc} at supercooled $T$). The configuration space is thus partitioned into the basins of attraction of the IS and the rate of {\it inherent structure transitions} (IST) may be measured. The IS-mapping is a powerful technique for the investigation of supercooled dynamics but it has not yet been exploited to anything like its full potential. Sastry et al. demonstrated \cite{sri} in the LJ mixture that the averaged IS energy, $<U_{is}(T)>$, undergoes a sharp drop (the \cite{angell} 'slope') from an upper high-$T$ to a lower low-$T$ plateau. They identify the beginning of the drop with $T_{A}$, and $T_{c}$ lies near the apparent bottom. The bottom energy is cooling-rate dependent and the true value, requiring cooling slow enough to maintain equilibration, remains unknown. It is appealing, within the landscape point of view, to relate the change in dynamical mechanism to occupation by the system of lower-lying IS. 

More direct investigations of the influence of the landscape on dynamics are possible. The minima of the squared potential gradient $W \equiv | \nabla U |^{2}$ include all the extrema, or critical points, of $U$, including minima of order $K=0$ and saddles of order $K > 0$. Minimizing $W$ defines a mapping to a configuration which need not be a minimum of $U$. With the {\it critical point mapping}, Cavagna, Angelani et al. \cite{andrea,ang} showed that minima dominate below $T_{c}$, saddles above. The IS-mapping assigns the system to a minimum even if it is in the upper reaches of the basin near the saddle-barriers; the critical point mapping does so only if it is actually close to the IS. They suggest \cite{andrea,ang} that the crossover is from motion among the minima (IS dynamics) at $T<T_{c}$ to motion among the saddles (saddle dynamics) at $T>T_{c}$. Either mapping may be used at any $T$. Reference to saddle dynamics or IS dynamics below indicates that it gives the simpler, more physical description. The {\it saddles ruled} \cite{andrea} regime is also one of {\it border dynamics}. Reaction coordinates leading to several different IS intersect at a multidimensional  or \cite{sw} 'monkey' saddle, so IS boundaries are closely spaced in the vicinity of a saddle. The system is always near a border, crossings are facile and frequent, and IST are not hops but merely unphysical 'bookkeeping' \cite{tkjc} events. A much better understanding of the $T>T_{c}$ region is being achieved with the new ideas about saddles and border dynamics.

Recently \cite{tkjc} we gave the first quantitative expression for $D$ in terms of the IST rate $<\omega_{is}>$ and the mean-square separation of successive IS. We argued that the long-time slope of the mean-square displacement per degree of freedom (MSD) of the IS-configuration, $R2(t) \equiv <(\Delta {\bf R}(t))^{2}>/6N$, is equal to $D$, as is so for the ordinary MSD, $r2(t)$. The IS-displacement is the sum of the IST-vectors (separations  of successive minima), $\Delta {\bf R}(t) = \sum_{\alpha=1}^{n(t)}{\bf \delta R}_{\alpha}$ after $n(t)$ transitions in time $t$; for any quantity $x$, the displacement $\Delta x(t) \equiv x(t) - x(0)$. A key quantity in this approach is \cite{tkjc} the IST-vector correlation, $C(\beta) \equiv <({\bf \delta R}_{\alpha} \cdot {\bf \delta R}_{\alpha+\beta}) \omega_{is}>$. In unit-density LJ, $N$ = 32 atoms, where $T_{c} = 0.52$ (LJ units), an IST-Markov approximation of random walking among the IS, predicts $D$ accurately for $T < T_{c}$ and substantially overestimates $D$ for $T > T_{c}$. The Markov approximation is equivalent to the statement that the IST-vectors are uncorrelated, $C(\beta)=0, \beta > 0$. Our result supports the idea of long sojourns in the basins below $T_{c}$, since the system then has time to lose correlation or memory between IST, leading to a random walk. Because of the uniquely weak $T$-dependence of $D$ at $T>T_{c}$ in LJ, 'transition to hopping' may actually be an accurate description of the crossover in this case. The overestimate of $D$ by the Markov approximation at $T>T_{c}$ is associated \cite{tkjc} with a long ranged anticorrelation of successive transition vectors, negative $C(\beta)$ at large $\beta$, necessary for frequent IST to result in relatively little displacement. {\it Anticorrelated IST-vectors are a signature of the 'bookkeeping' IST occuring in border dynamics.}

A problem arises in trying to learn about landscape dynamics with the IS-mapping. In the absence of long-range correlations the configuration space of the liquid should be regarded as a composite of the spaces of $N_{crr}$ independent local regions of $z^{*} \sim O(1)$ particles. A simple model \cite{bhup} uses 3$N$ sinusoidal potentials. Local regions and Adam-Gibbs \cite{ag} CRR need not be equivalent, but we will treat them as such. With independent IST occurring in each region, and with a transition for the whole system recorded when any region changes, $<\omega_{is}> \sim O(N)$ and the distribution of waiting times between IST, $g(t)$, is correspondingly \cite{sw}, and unphysically, skewed towards short time $\sim O(1/N)$. Computational constraints then indicate that one should simulate the smallest realistic system; with sufficiently large $N$ an IST will be observed on every time step and no meaningful calculation will be possible. More fundamentally, the desired properties are those of the CRR, the building blocks of the macroscopic liquid, but straightforward simulation yields those of the homogenized composite. With increasing $N$ the former become more difficult to disentangle from the latter. This is also true for static quantities. For example, whatever the distribution of $U_{is}$ in a local region, it will be Gaussian for the entire system at large $N$. Even for the best choice of $N$ some new theory is required to interpret the simulation data.

In this article we further examine the $T$-dependent mechanism of diffusion in unit-density LJ, $N$ = 32, via three new landscape dynamical entities, and with an initial attempt to derive local relaxation from simulated composite dynamics. The first is $r2_{is}(t) \equiv <(\Delta {\bf r}(t))_{is}^{2}>/6N$, the MSD {\it with no contribution from IST}. Starting in a thermal ensemble of basins, $r2_{is}(t)$ is calculated from only those trajectories which remain in the initial basin at time $t$. The second is $R2(t)$ and the third is the distribution $g_{loc}(t)$ of waiting times in an IS {\it for a local region or CRR}.  We present a simple method to obtain $g_{loc}(t)$ from $g(t)$ and $N_{crr}$, estimated from the averaged participation ratio, $<Pr>$, of the IST-vectors. The number of degrees of freedom (not particles) participating in an IST is roughly $<Pr>$; thus $z^{*} \approx <Pr>/3$, $N_{crr} \approx 3N/<Pr>$.

Comparison of $r2_{is}(t)$, $R2(t)$, $r2(t)$ and $g_{loc}(t)$ casts considerable light upon the relation of diffusion to the landscape. For $t >> \tau_{pl}$, where $\tau_{pl}$ is a characteristic plateau time, $r2_{is}(t)$ attains a constant value, due to \cite{lav} the finite volume of the basin. This is true even at high $T$ where the system moves freely among the IS, since only trajectories which remain in the original IS contribute (thus averaging is difficult). At $T$ = 1.10 the plateau is reached quickly but a small interval of  linear $t$-dependence is visible at intermediate times, from which an {\it intrabasin diffusion coefficient}, $D_{is}$, may be calculated. Below $T_{A} \approx 1.0$ the diffusive interval in $r2_{is}(t)$ expands; $T_{A}$ is estimated as the top of the slope in $<U_{is}(T)>$. More cooling reveals that the $T \sim T_{c}$ regime is {\it intrabasin dominated} with $D = D_{is}$ - the true diffusion constant may be calculated in a single basin! Thus, as for high $T$, diffusion cannot consist of a sequence of activated IST hops, but the mechanism is quite different from the high-$T$ mechanism. Intrabasin diffusion arises from the roughness of the landscape at the upper elevations of a basin and provides a mechanism for confinement different from inter-basin barriers but nonetheless capable of producing \cite{tkjc} a Markov chain of IST under the right conditions. We suggest that the elementary steps in this process are transitions among the domains of the saddles connected to the IS, ie, intrabasin diffusion is saddle dynamics. Intrabasin {\it dominated} diffusion is also IS dynamics, since the motion among the IS is then a simple random walk; at intermediate $T$ both descriptions are appropriate.

The system does not sense the finite size of the basin until the mean local waiting time, $\tau_{loc}$, reaches $\tau_{pl}$. We thus have a new criterion for the onset of the low-$T$ IS-dynamics mechanism and, indeed, it is found that $\tau_{loc}$ crosses $\tau_{pl}$ slightly below $T_{c}$. The analysis would make no sense with an $N$-dependent mean time taken from $g(t)$. More information is obtained by comparison of $R2(t)$ with $r2(t)$. Let ${\bf r}(t) = {\bf R}(t) + {\bf u}(t)$, where \cite{lav} ${\bf u}(t)$ is the return vector, the separation of the system from its IS. If $\Delta {\bf R}$ and $\Delta {\bf u}$ are uncorrelated, $r2(t)$ must lie above $R2(t)$. This is so at low $T$ but as $T$ increases $R2(t)$ rises above $r2(t)$. The relation of $R2(t)$ to $r2(t)$, indicating the correlation of  $\Delta {\bf R}$ and $\Delta {\bf u}$, yields a {\it border index} $B(T)$ which complements the information contained in $C(\beta)$ regarding IS vs saddle dynamics. In sum, with some novel applications of the IS- mapping we achieve an extremely detailed description of the $T$-dependent mechanism of diffusion.
\section{Simulation Methods \label{sim}}
Molecular dynamics simulations are performed on unit-density supercooled LJ \cite{tkjc,tk,jctk}, $N$ = 32, with the methods described in ref~\cite{tkjc}. As discussed there, and above, a small system should be used for IST dynamics. Increasing $N$ beyond the number required to 'solvate' a CRR merely creates an intractably large IST rate and makes it difficult to extract the properties of a CRR from those of the composite landscape. Is $N$=32 large enough? The diffusion constants are close to those for \cite{jctk} $N$=108 and \cite{tk} $N$ = 256, with slightly weaker $T$-dependence. The crystal melts at $T_{m} \sim 1.6$ and \cite{tkjc} fitting $D(T)$ yields $T_{c} = 0.52$, while at $N$=256 \cite{tk} $T_{m} \sim 1.8$ and (for 'modified' LJ) \cite{ang} $T_{c} = 0.475$. The saddle order \cite{andrea,ang} $K(T)$ extrapolates to zero at \cite{jctk2} $T_{c}$ (from $D(T)$ by definition), as is so for \cite{jctk} $N$ = 108 and \cite{ang} for modified LJ, $N$ = 256. The plot \cite{tkjc} of $<U_{is}(T)>$ is similar to that of ref~\cite{ang}, reaching the 'bottom' of the landscape at the same $T \approx 0.50$, and with the gentler drop from the high-$T$ plateau expected for small $N$. Clearly, the essential physical features of the crossover - we make no claim about deeply supercooled liquids -  are  present at $N$=32 and using a larger system would simply fine-tune various numerical estimates at enormous \cite{tkjc} computational cost.

Natural LJ units will be used throughout. Rather than making a single long MD run in the supercooled liquid, we average all calculated quantities over an ensemble of quenches, starting from different high $T$ = 5.00 configurations, to avoid \cite{jwalk} broken ergodicity. The hot liquid is cooled in one step to a temperature in the 1.20-0.60 range. The system is equilibrated for 2.5 $\tau_{LJ}$, data are gathered for 62.5 $\tau_{LJ}$, $T$ is decreased by 0.02 and the process is repeated 10-25 times, generating a single quench run; most quenches sampled 16 $T$. The cooling rate is $3.08X10^{-4}$. Sastry et al. \cite{sri} found in an LJ mixture that quenches with a cooling rate of $2.70X10^{-4}$, close to ours, exhibited $\sim 75 \%$ of the fall in $<U_{is}(T)>$ attained by quenching almost 100 times slower and reached an apparent bottom somewhat below $T_{c}$; this should be adequate for probing the crossover.  At $N$ = 32 the abrupt decreases in $U$ signaling solidification, common at $N$ = 256, do not occur but some quenches develop solid-like pair distributions and these are discarded. Results are averaged over 23 quenches at the lowest $T$ and 30 at the highest. Quench-to-quench fluctuations are much larger than any systematic changes over $T$ = 0.02, so we also average results at each $T$ with those from the next higher and lower $T$. Even so our data are somewhat noisy but the trends are clear for $1.10 \geq T \geq 0.34$, encompassing the crossover range $T_{A}(\approx 1.0) \geq T \geq T_{c}(= 0.52)$. We believe that our results are well averaged and equilibrated down to $\sim T_{c}$, and are out of equilibrium (but 'well averaged' over the set of non-equilibrium configurations allowed by the quench) at the lowest $T$.

Conjugate gradient minimizations are performed every 5 time steps  ($dt$ = .00125), or 160 minimizations/$\tau_{LJ}$. Since the range of $<\omega_{is}(T)>$ is from 9.0 IST/$\tau_{LJ}$ at $T$ = 1.10 to 0.23 at $T$ = 0.34, this is sufficient. The equivalence of conjugate-gradient and steepest-descent for the relevant $T$ was verified. The distribution of IST-vector (logarithmic) lengths $d$ is bimodal, and \cite{donati} to avoid counting contributions of two-level systems and anharmonicities irrelevant to diffusion we record a transition for $d$ in the large-displacement lobe only. Similar considerations have \cite{inm} entered the efforts to relate $D$ to $Im-\omega$ instantaneous normal modes, where {\it non-diffusive} modes must be discarded.

At each $T$ we obtain the three MSD, $r2_{is}(t)$, $R2(t)$, and $r2(t)$ (yielding $D$), the distribution of waiting times, $g(t)$, and the averaged participation ratio $<Pr>$ of the IST-vectors. With the $t$-dependent IS in hand evaluation of $R2(t)$ and $<Pr>$ is straightforward. The intrabasin MSD $r2_{is}(t)$ is found from an ordinary MSD algorithm with the condition that, when an IST is detected, the current run is terminated and a new run is begun (new origin of coordinates, MSD=0). Binning the times spent in an IS, instead of simply calculating the averaged IST rate, is all that is required for $g(t)$.

\section{Landscape Dynamical Properties \label{scape}}
\subsection{Intra- and Inter- Basin Dynamics \label{scape1}}
Distributions of IS waiting times $g(t)$ are described very well by a KWW (stretched exponential) function $exp(-(t/\tau)^{\beta})$ over the entire temperature range. The KWW fit is superior to a sum of two exponentials, even though the latter has one more adjustable parameter. The average $\beta$ = 0.50, and deviations appear to be noise with no systematic $T$-dependence. To investigate trends in $\beta$ over a broader $T$-range we obtained data at normal liquid $T$ = 2.00 and also found, to our surprise, $\beta$ = 0.50.

One hesitates to read very much into the intriguing half-integral value of $\beta$ because $g(t)$ is a composite property, dependent upon $N_{crr}$. An estimate of $g_{loc}(t)$ may be obtained as follows. The probability that there is no IST in time $t$ is $P_{no}(t) = \int_{t}^{\infty}dt'g(t')$. In a composite, no transition means that no CRR has a transition, and $P_{no}(t) = (\int_{t}^{\infty}dt'g_{loc}(t'))^{N_{crr}}$. Equating the two expressions, solving for $\int_{t}^{\infty} dt' g_{loc}(t')$, and taking $d/dt$ yields
\begin{equation}
g_{loc}(t) = \frac{g(t)}{N_{crr}(\int_{t}^{\infty} dt' g(t'))^{(1-1/N_{crr})}}.
\label{gloc}
\end{equation}

\begin{figure}
\psfig{figure=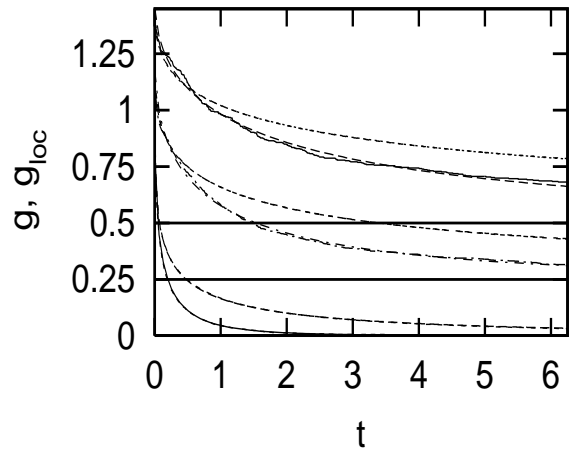,height=2.75in,width=3.25in,angle=-90}
 \caption{ 'Composite' and local waiting time distribution pairs $g(t)$ and $g_{loc}(t)$ with overlapping KWW fits at $T$ = 0.34, 0.50 and 1.00, top to bottom. Curves are set to unity at $t$ = 0, shifted for display, and decay to zero at the horizontal lines; local distribution has slower decay.} \label{ggloc}
 \end{figure}

\begin{figure}
\psfig{figure=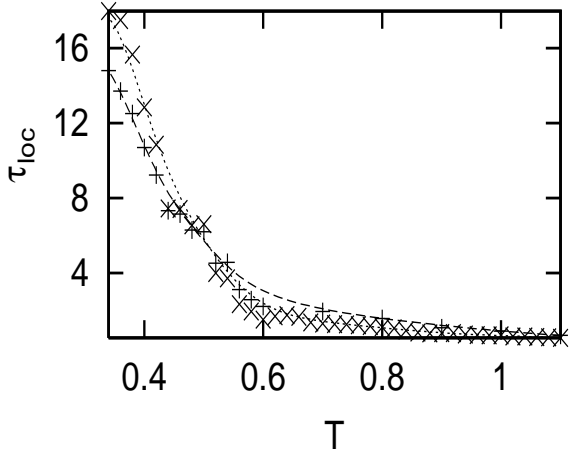,height=2.75in,width=3.25in,angle=-90}
 \caption{ $T$-dependence of local waiting time from $g_{loc}$ (+) and from $N_{crr}/ <\omega_{is}>$($\times$); smooth curves are spline fits.} \label{tau}
 \end{figure}

The averaged participation ratio of the IST-vectors is nearly constant, fluctuating between 19-22 over the $T$-range with an average of 21 for $z^{*}$=7.0 atoms in a CRR and $N_{crr}$ = 4.6 CRR in the simulation box. Using the average $N_{crr}$ at all $T$ we calculate $g_{loc}$ which are also well described by a KWW form, with $T$-independent $\beta$ = 0.36. Some representative $g_{loc}(t)$ from Eq~\ref{gloc}, simulated $g(t)$ and fits with $\beta$ = 0.36 and $\beta$ = 0.50, respectively, are presented in Fig~\ref{ggloc}. The shift of $g_{loc}(t)$ to longer times is
\noindent evident. Processes with KWW distributions are not well characterized by a single time, but the best compromise is the correlation time, $\tau_{loc} = \int_{0}^{\infty} dt g_{loc}(t)/g_{loc}(0)$. Fig~\ref{tau} provides further evidence \cite{tkjc,jctk,ang} that the low-$T$ diffusive mechanism sets in at $T \sim 0.50$, where $\tau_{loc}(T)$ begins to increase strongly. Also shown is $N_{crr}/<\omega_{is}>$, the obvious intensive 'composite corrected' time available from the average IST rate, in reasonable agreement with $\tau_{loc}(T)$.

\begin{figure}
\psfig{figure=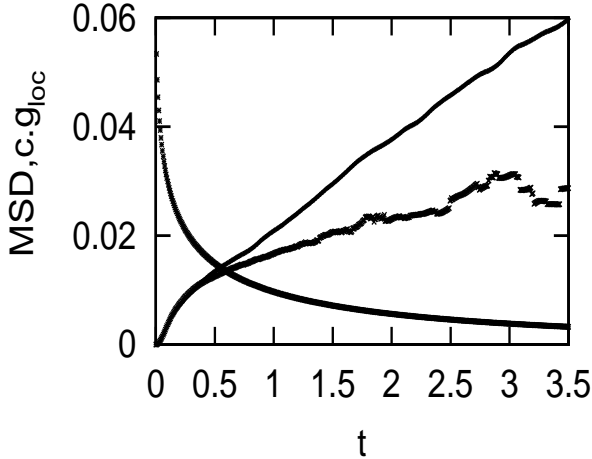,height=2.75in,width=3.25in,angle=-90}
 \caption{ Conventional (upper) and intrabasin MSD at $T$ = 1.10. Decaying curve is distribution of local waiting times, scaled to cross intra-MSD at $\tau_{loc}$. Plots are closely-spaced data points.} \label{msd110}
 \end{figure}

Let us now view the changes in the mechanism of diffusion with decreasing $T$ through the paired conventional and intra-IS MSD, $r2(t)$ and $r2_{is}(t)$, combined with our knowledge of $\tau_{loc}(T)$. The $T \sim T_{A}$ scenario is found at $T$ = 1.10, Fig~\ref{msd110}, a relatively high (although supercooled) temperature $\sim 2T_{c}$ where thermal energy is sufficient for the system to move freely among closely-spaced IS
\noindent borders. The unconstrained MSD diverges from $r2_{is}(t)$ at $t \sim 0.36$, before it has even fully established its linear diffusive form. For the same reason the trajectories which stay in the basin explore it quickly; the short time rise in $r2_{is}(t)$ bends over to form the plateau at about $\tau_{pl} \sim 2.0$ with only a glimpse of intermediate-$t$ behavior. Even so, almost all trajectories leave the basin before its finite size is felt, with $\tau_{loc}=0.68<<\tau_{pl}$ (thus $r2_{is}(t)$ is noisy at long time). This is illustrated with the inclusion of $g_{loc}$, scaled to intersect $r2_{is}(t)$ at $\tau_{loc}$. The IST are 'bookkeeping' events as border dynamics prevails, $D$ is not \cite{tkjc} proportional to $<\omega_{is}>$, there is no effective barrier to leaving a basin and hopping is surely not an apt description. One might say either that intrabasin dynamics are irrelevant to diffusion, or that no distinction exists between inter- and intra- basin dynamics.

\begin{figure}
\psfig{figure=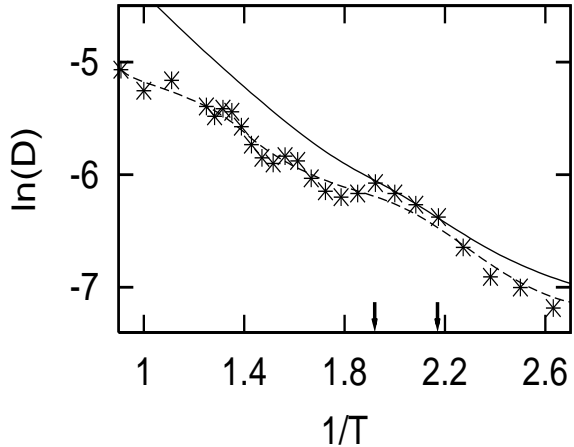,height=2.75in,width=3.25in,angle=-90}
 \caption{ Natural logarithms of intra-basin diffusion constant $D_{is}$ (dashed line spline fit and points) and true $D$ (solid line) vs $1/T$. {\it Intrabasin dominated} $D = D_{is}$ range delimited by arrows, left arrow coincides with $T_{c}$.} \label{ddis}
 \end{figure}

At $T = 0.80 < T_{A}$, $r2_{is}(t)$ looks rather different - the difusive interval has grown to the point that it resembles a conventional MSD and easily allows estimation of an intrabasin $D_{is}$. The plateau must be reached eventually, but in the current simulation all we can say is $\tau_{pl} > 6.25$. Comparison with $r2(t)$ shows (Fig~\ref{ddis}) that $D_{is}<D$, with $D_{is}$=0.0045, $D$=0.00714. The two MSD have a brief interval of overlap at the beginning of their linear regimes but soon diverge, so the mechanism is not governed by
\begin{figure}
\psfig{figure=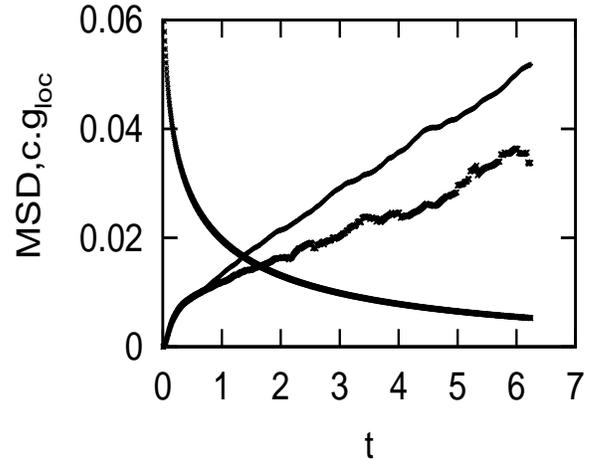,height=2.75in,width=3.25in,angle=-90}
 \caption{ As Fig 3 for $T$ = 0.80.} \label{msd080}
 \end{figure}
\noindent  intrabasin diffusion and is probably unchanged from $T$ = 1.10. Nonetheless we believe that the emergence of intrabasin diffusion is significant and characteristic of the upper crossover regime. Basins are not harmonic away from the minima and apparently below $T_{A}$ the higher elevations are rough enough to produce diffusion. An equivalent statement is that the process is best viewed as a random walk among the saddles \cite{andrea,ang}. Configurations in the upper part of a basin are closer to the saddles than to the IS and will be assigned accordingly by the critical point mapping. A random walk among saddles connected to a basin will appear as intrabasin diffusion. At this $T$ ordinary diffusion is also saddle dynamics, so with $D_{is}<D$ it follows that unconstrained saddle dynamics is faster than intrabasin saddle dynamics. The need to utilize an inefficient random walk to escape the basin, and of course the lower thermal energy, has caused the plateau time to increase significantly from its $T$ = 1.10 value and $\tau_{loc}<<\tau_{pl}$ holds even more strongly despite an increase in $\tau_{loc}$.

\begin{figure}
\psfig{figure=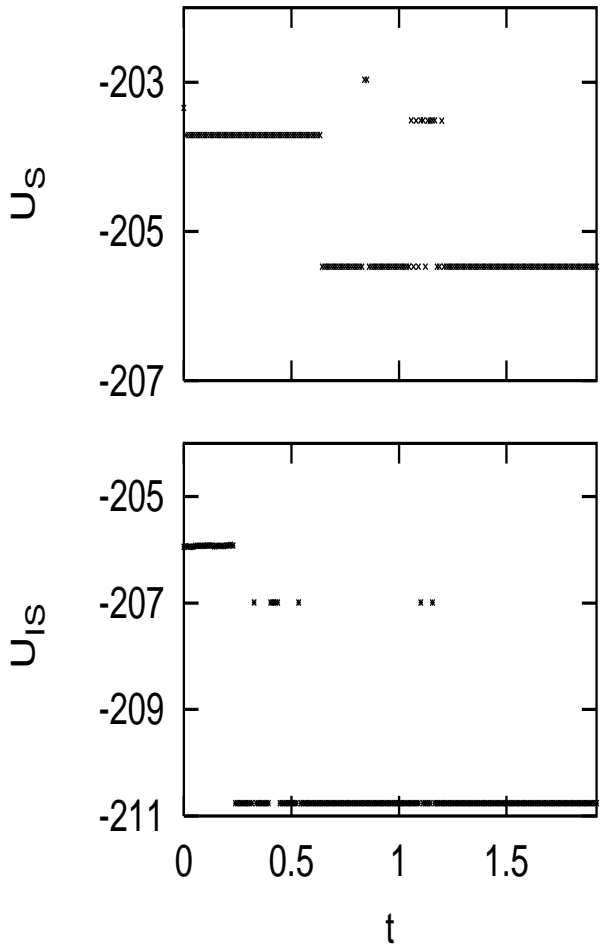,height=5.5in,width=3.25in,angle=-90}
 \caption{Saddle (upper) and IS energies vs $t$, $T$ = 0.70.} \label{sadIS}
 \end{figure}

In our picture of intrabasin diffusion, saddle transitions occur while the IS does not change. We previously discussed 'bookkeeping' IST, where the IS changes but there is little actual motion. With several IS available from a saddle, motion about the basin of attraction of a single saddle naturally generates bookkeeping IST; the IS changes but the saddle does not.  Both possibilities, which coexist at intermediate $T$,  may be observed with a time-dependent critical point mapping, made possible by \cite{jctk2} a new, very efficient algorithm. Fig~\ref{sadIS} shows IS
\noindent and saddle energies vs $t$ at $T=0.70$. In the time interval $0.00 \leq t \leq 0.70$ the system undergoes seven IS transitions while assigned to the same saddle (bookkeeping IST). For $0.70 \leq t \leq 1.15$ the IS is constant and there are two saddle transitions at $t \sim 0.8$, followed by rapid exchange between a pair of saddles starting just after $t = 1.00$ (intrabasin diffusion).

\begin{figure}
\psfig{figure=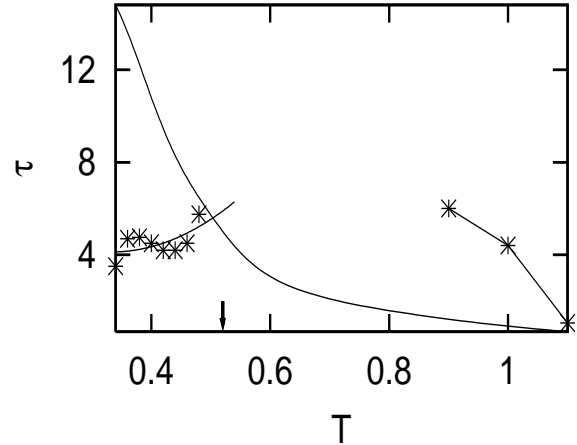,height=2.75in,width=3.25in,angle=-90}
 \caption{ Local waiting time $\tau_{loc}$ and, in two segments (line plus $\star$), rough estimate of plateau time $\tau_{pl}$ vs $T$; low-$T$ data shown with quadratic fit. Arrow indicates $T_{c}$.} \label{ttpl}
 \end{figure}

Cooling to $T$ = 0.50 $\sim T_{c}$ produces a remarkable result.
The period of overlapping linear $t$-dependence of $r2_{is}$ and $r2$ is now substantial, and (Fig~\ref{ddis}) {\bf the true diffusion constant may be calculated from trajectories with no IST.} The system now stays in a  basin long enough
that on average intrabasin diffusion is fully developed before an IST occurs, and before the plateau is reached (although $\tau_{loc}$ is now approaching $\tau_{pl}$, Fig~\ref{ttpl}). Under these circumstances diffusion is {\it intrabasin dominated}, and the IST simply act to keep the process going, ie, to avoid the influence of finite basin size. Intrabasin and unconstrained saddle dynamics (ordinary diffusion) are indistinguishable. We thus have some new
\begin{figure}
\psfig{figure=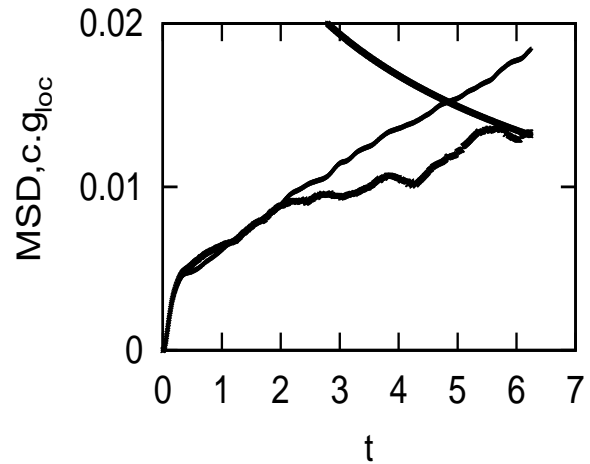,height=2.75in,width=3.25in,angle=-90}
 \caption{ As Fig 3 for $T$ = 0.50.} \label{msd050}
 \end{figure}
\noindent 
perspective on the changes around $T_{c}$. There the chain of IST becomes \cite{tkjc} a Markov process, and an obvious explanation is that sufficiently long confinement of the system in a basin allows successive IST to lose correlation. While conventionally confinement arises from energy barriers, we find that the system diffuses to the border and crosses with ease. Confinement is produced by diffusion itself; the time required to escape by the random walk is long even absent a barrier at the border. A sufficiently slow walking about the saddles connected to a basin can produce a Markov process among the IS themselves, and IS dynamics is also a good description of the intrabasin dominated regime. Upon close examination, the crossover is not simply from saddle dynamics at high $T$ to IS dynamics at low $T$. There is a third, intermediate-$T$ region where both descriptions are correct, defined by the condition $D=D_{is}$.

\begin{figure}
\psfig{figure=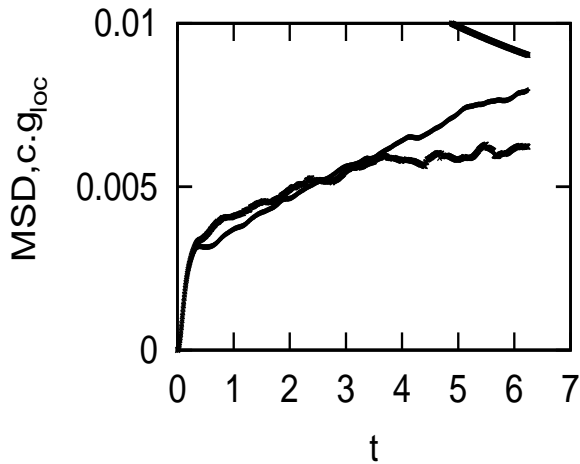,height=2.75in,width=3.25in,angle=-90}
\caption{ As Fig 3 for $T$ = 0.34.} \label{msd034}
\end{figure}

Arrhenius plots of  $D_{is}$ and $D$ are displayed in Fig~\ref{ddis}. Estimates of $D_{is}$ are possible for $1.10 \geq T \geq 0.38$. Linear behavior of $r2_{is}(t)$ is not cleanly visible below $T$ = 0.38 and is marginal at $T$ = 1.10; we expect it disappears at slightly higher $T$, but collecting data where trajectories leave a basin so quickly is difficult. The intrabasin dominated $D=D_{is}$ range, marked by arrows, is approximately $T_{c}=0.52 \geq T \geq 0.46$. Starting from high $T$, the upper limit of intrabasin dominance is reached when two conditions hold: $T$ is low enough for $r2_{is}(t)$ to have a diffusive $t$-interval, and $\tau_{loc}$ adequately exceeds the time at which that interval begins. For the lower limit, either of two conditions may be true: $\tau_{loc}$ substantially exceeds $\tau_{pl}$, so a barrier to leaving the basin exists and activation becomes important, or the linear portion of $r2_{is}(t)$ shrinks so that it no longer represents the intrabasin motion. In fact, both appear to set in gradually and simultaneously.
\noindent  At our lowest temperature, $T$ = 0.34 (Fig~\ref{msd034}), $r2_{is}(t)$
\noindent rises quickly to the plateau just as at the high $T$ = 1.10; a linear region of this curve cannot be identified and $\tau_{pl}$ is now back within the 6.25 $\tau_{LJ}$ window.

Quantitative estimates of $\tau_{pl}(T)$, along with $\tau_{loc}(T)$, are shown in Fig~\ref{ttpl}; at intermediate $T$ all we can say is
\noindent  $\tau_{pl}>6.25$. While $\tau_{pl}$ is short at high $T$ because of the abundant thermal energy, it is short at low $T$ because the system occupies the lower regions of the basin only. Thus a) the plateau is reached after a shorter displacement, and b) the intra-basin motion is mostly fast harmonic oscillations, not diffusion. The $\tau_{pl}$  and $\tau_{loc}$ curves cross
at $T \approx 0.51$, and we propose $\tau_{pl} = \tau_{loc}$ as a new way to define a crossover temperature. Below $T$  =  0.51 the system explores the accessible portion of the basin more quickly but exits more slowly, indicating barriers at the border. Gradually saddle dynamics loses out to IS dynamics as the physical description, $D \neq D_{is}$ below $T$ = 0.46, and barrier crossings assume the rate-limiting role for diffusion.

In addition to the standard analysis of $D(T)$ and the results \cite{ang,tkjc,jctk} of very recent work, we now have two new indicators  pointing to a change in the mechanism of diffusion at $T \sim 0.50$. The local waiting time $\tau_{loc}$ crosses $\tau_{pl}$ at $T$ = 0.51, and intrabasin dominance sets in at $T$ = 0.52 with $D = D_{is}$ for $0.52 \geq T \geq 0.46$. However the crossover develops over a wide temperature range, and involves much more than the presence or absence of hopping, as explained in the last few paragraphs. We hope it is clear that the IS-mapping is a powerful tool for studying the dynamical crossover.

\subsection{Landscape Vector Correlations, Saddle Dynamics and Borderism \label{scape2}}
IST-vectors are anticorrelated \cite{tkjc} at high $T$, and correlation decreases through the crossover range until the IST-Markov approximation becomes accurate below $T_{c}$. A substantial negative value of $C(\beta)$ is a sign of saddle or border dynamics, while $C(\beta) \sim 0$ indicates IS or hopping dynamics. IS borders are closely spaced near a multidimensional saddle. Anticorrelation is required for the 'bookkeeping' IST arising from a small motion through such a region to not, wrongly, predict a large motion.  It implies that saddle dynamics is a more physical description than IS dynamics. A simple process appears complicated when formulated with inappropriate elementary steps, and loss of correlation signals the onset of the low-$T$ mechanism, which is truly IS dynamics. 

\begin{figure}
\psfig{figure=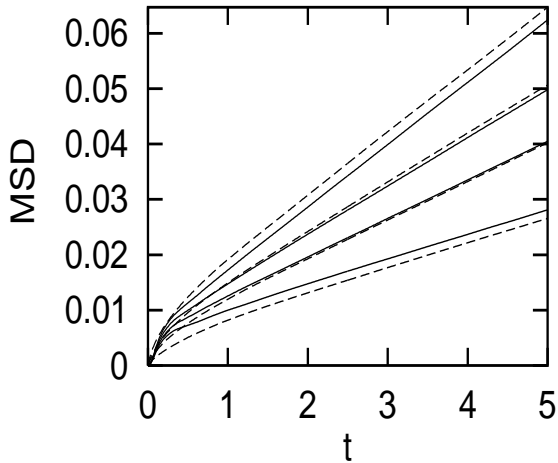,height=2.75in,width=3.25in,angle=-90}
 \caption {Ordinary MSD (solid) and IS-MSD (dashed) vs $t$; top to bottom, $T$ = 1.00, 0.90, 0.80, 0.70.} \label{msds}
 \end{figure}

Analysis of the MSD within the basin of attraction of an IS, $r2_{is}(t)$, proved quite fruitful. We will now see that such is also the case for the MSD of the IS itself, $R2(t)$, leading to another informative correlation, that of the return vector and the IS configuration. With the definitions from the Introduction,
\begin{eqnarray}
r2(t) = R2(t) + <(\Delta {\bf u}(t))^{2}>/6N + \nonumber \\
< \Delta {\bf R}(t) \cdot \Delta {\bf u}(t)>/6N.
\label{msdeq}
\end{eqnarray}
It follows that $r2(t) \geq R2(t)$ unless the IS displacement, $\Delta {\bf R}(t)$, and the return vector displacement, $\Delta {\bf u}(t)$, are anticorrelated. Fig~\ref{msds} shows that at $T$ = 1.00, $R2(t)$ lies 
\noindent {\it above} $r2(t)$, they nearly sumerimpose at $T$ = 0.80, and the 'obvious' $r2(t) \geq R2(t)$ holds at  $T<0.80$. Thus the pattern of anticorrelation of IST-vectors $\delta {\bf R}_{\alpha} \cdot \delta {\bf R}_{\alpha + \beta}$ at high $T$, decreasing  with decreasing $T$, is repeated with the $\Delta {\bf R}(t) \cdot \Delta {\bf u}(t)$ correlation.

\begin{figure}
\psfig{figure=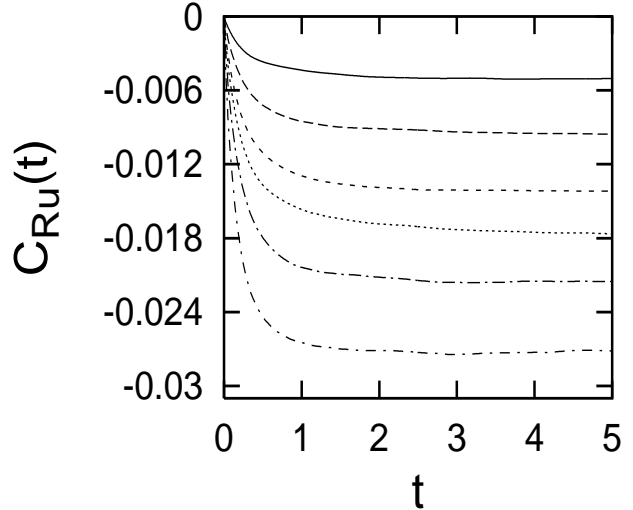,height=2.75in,width=3.5in,angle=-90}
\vspace{.25in}
 \caption {Correlation $C_{Ru}(t)$, amplitude increasing with increasing $T$; top to bottom, $T$ = 0.60, 0.70, 0.80, 0.90, 1.00, 1.20.}
 \label{CRu}
 \end{figure}

For a more quantitative analysis we have calculated $<(\Delta {\bf u}(t))^{2}>$. One anticipates a rapid decay of $<{\bf u}(t) \cdot {\bf u}(0)>$ leading to $ <(\Delta {\bf u}(t))^{2}> \rightarrow 2<q^{2}>$ and this expectation is correct. The correlation $C_{\bf Ru}(t) \equiv < \Delta {\bf R}(t) \cdot \Delta {\bf u}(t)>/6N$, determined from $r2(t)$, $R2(t)$, and $<(\Delta {\bf u}(t))^{2}>$ via Eq~\ref{msdeq}, decays quickly (Fig~\ref{CRu}) to an asymptotic negative value which decreases in amplitude with decreasing $T$. We suggest that $C_{\bf Ru}$ is a direct measure of the degree of {\it borderism} - the extent to which the system 'lives' on the IS-borders \cite{kurchan}.

\begin{figure}
\psfig{figure=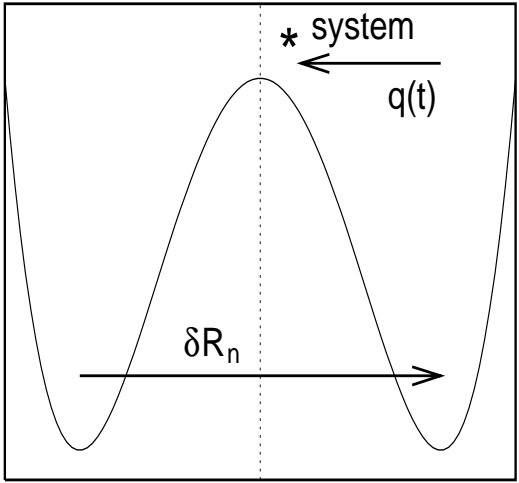,height=2.75in,width=3.25in,angle=-90}
 \caption{ Return vector ${\bf u}(t)$ and IST-vector $\delta {\bf R}_{n}$ at time $t$ just after $n$'th (most recent) IST.}
 \label{anticor}
 \end{figure}

Expressing $\Delta {\bf R}(t)$ as the sum of the IST-vectors the most important contributions are evident,
\begin{eqnarray}
C_{\bf Ru}(t) =  <\delta {\bf R}_{n} \cdot {\bf u}(t)> - <\delta {\bf R}_{1} \cdot {\bf u}(0)> \nonumber \\
+ (smaller \: terms).
\label{Crq}
\end{eqnarray}
The first term on the RHS may be understood with Fig~\ref{anticor}, a one-dimensional illustration of the situation at time $t$, just after the $n$'th IST. The system has moved from the left- to the right-hand basin and it follows that $\delta {\bf R}_{n}$ and ${\bf u}(t)$ point in opposite directions, with a negative contribution to $C_{\bf Ru}$. For the second term on the RHS, set $t$ = 0 in Fig~\ref{anticor}. The first IST will occur from right to left in the future so change $\delta {\bf R}_{n}$ to $\delta {\bf R}_{1}$ and reverse its direction. The vectors now point in the same direction so with the minus sign in Eq~\ref{Crq} this contribution to $C_{\bf Ru}$ is also negative. In the $3N$ dimensional configuration space (or $3N_{crr}$ dimensional local region space) the vectors will be distributed about the $d$ = 1 relative orientations, reducing the dot products.

\begin{figure}
\psfig{figure=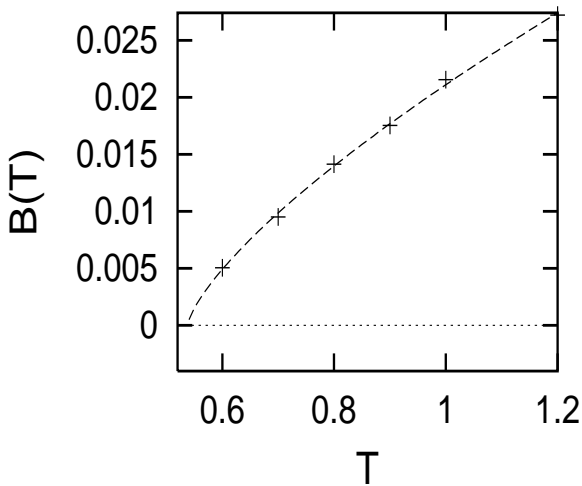,height=2.75in,width=3.25in,angle=-90}
\vspace{.2in}
 \caption{ Border index $B(T)$ (points) and power law fit, tending to zero at $T$ = 0.53.}
 \label{bordr}
 \end{figure}

The geometrical arguments apply when ${\bf u}(0)$ is evaluated just before, and ${\bf u}(t)$  just after, an IST. For border dynamics these conditions 'always' hold and $C_{\bf Ru} < 0$. On the other hand if the system oscillates about the IS for long periods between IST, the low-$T$ scenario, the return vector has no special relation to any IST vector at any time and $C_{\bf Ru} \sim 0$. The correlation of  $\Delta {\bf R}(t)$ and  $\Delta {\bf u}(t)$ is a maximum for pure border dynamics and vanishes in a hopping model. The interesting part of $C_{\bf Ru}$ is its asymptotic value and we define
\begin{equation}
B(T) \equiv -C_{\bf Ru}(\infty,T)
\end{equation}
as the {\it border index}. The $T$-dependence of the border index is shown in Fig~\ref{bordr}. It 
extrapolates to zero via
\noindent power-law fit for $1.2 \geq T \geq 0.60$ at $T$ = 0.53, yielding another physically appealing route to a crossover temperature. The fit range differs from that used for $T_{c}$ but $B(T)$  is close enough (power = 0.74) to linear in $T$ that we expect only a small effect.

\section{Discussion \label{disc}}
In the foregoing we have examined the $T$-dependent physical mechanism of diffusion in unit-density LJ. For some perspective, it is useful to ask why this is a nontrivial problem. LJ is classified \cite{dk} as weak/nonfragile, indicating weak $T$-dependence at high $T$ and a constant activation energy at low $T$. The simulated diffusion constant is indeed well represented by
\begin{equation}
 D(T)=0.39Texp(-1.16/T)
\label{fit}
\end{equation}
for $1.10 \geq T \geq 0.34$; $E_{A}$ = 1.16. The factor of $T$ is appropriate because $D \sim T$ at high $T$ and constant density. Linear or other multiplicative powers are not very important when fitting over several decades but we have only $20X$ and they significantly influence the goodness of fit and the value obtained for the activation energy. Despite the simplicity of Eq~\ref{fit}, fitting \cite{jctk2} the same data to a power law for $2.0 \geq T \geq 0.60$ yields a reasonable \cite{ang} $T_{c}$.  One might well wonder if anything is going on beyond a gradual change from $T>>E_{A}$ to $T<<E_{A}$. That is a bare-bones prescription for a 'transition to hopping', consistent with the IST approaching a Markov chain \cite{tkjc} and the saddle order extrapolating to zero \cite{jctk,jctk2}. 

However it then seems odd that $E_{A}$ = 1.16 does not appear as a characteristic dynamical temperature instead of various $T \sim 0.50$, and the sharp drop in $<U_{is}(T)>$ around $T_{c}$ suggests that the crossover is more interesting. To investigate this crucial point we have begun \cite {jctk2} to catalog the saddles and barrier heights in unit-density LJ. Starting from the thermal configuration the IS is determined and the connected first-order saddles are found \cite{jctk2} by eigenmode following. Steepest descent from the saddle yields barrier heights $\Delta U$, which may be collected as a function of IS energy \cite{andrea2} or of the original temperature. Preliminary results \cite{jctk2} are that the barrier heights increase through the slope region \cite{angell,sri} of $<U_{is}(T)>$ from $<\Delta U(T = 1.20)>$ = 2.8 to $<\Delta U(T = 0.40)>$ = 3.7; $T$ is less than the averaged barrier height for the entire range of this study. The same result has recently been obtained \cite{andrea2} in a soft-sphere mixture. Comparisons of barrier heights to $T$ must be made carefully. Most discussions of barrier crossing assume that $\Delta U$ is much greater then $T$, but that does not apply here and one might argue that the relevant height is $(\Delta U-T)$. Even with such a modification, average barrier heights remain larger than $E_{A}$, indicating - along with the new results found above  - that the mechanism of diffusion in LJ is highly non-trivial.

There are two explanations for the clear presence of a high-$T$ mechanism with free motion among the basins at $T > T_{c}$ despite an apparent requirement for activated hopping.  One is the broad distribution of barrier heights, reflected in the low $\beta = 0.36$ for the KWW distribution of local waiting times. Even if $T$ falls below $<\Delta U(T)>$, there will be no transition to low-$T$ dynamics if enough low barriers remain. The average alone does not adequately describe the physics. Furthermore there are several ways to perform the average \cite{jctk2} and a physically motivated choice must be made. For example a calculation of $<\Delta U(T)>$ including saddles of all orders would be nonsense, since high barriers which are never visited in thermal motion would dominate. Averaging over first order saddles eliminates many irrelevant high barriers, but not necessarily all. Other topological features such as the connectivity of the IS and saddles should be important as well. The landscape determines $T_{c}$ and more generally $D(T)$ through the interplay of several physically significant effects. Note that one important landscape quantity, $<U_{is}(T)>$, is a 'thermodynamic' property of the basins. One can imagine different liquids with the same $<U_{is}(T)>$ and different T-dependent distributions of barrier heights, giving rise to quite different activation energies and other aspects of dynamics. 

Another is found in Cavagna's \cite{andrea,andrea2} argument that barrriers are irrelevant while the saddle mechanism dominates, so the action of a high-$T$ mechanism while $T$ is less than $<\Delta U>$ need present no conundrum. Then, when the saddle mechanism shuts down and activation becomes required at $T \sim 0.50$, the barrier heights have already completed most of their growth. The system will exhibit approximate Arrhenius behavior upon further cooling. Cavagna gives a 'fragile' scenario where the barriers are already $>> T$ at $T_{c}$, for a very strong increase in relaxation time. However although we have  $<\Delta U(T_{c})>/T_{c} \sim 6$, $E_{A}$ is only $\sim 2T_{c}$, presumable due to the importance of low-barrier paths. Relatively weak Arrhenius behavior at $T \leq T_{c}$ results, consistent with LJ being \cite{dk} a weak/nonfragile liquid.

Even if $E_{A}=1.16$ is not meaningful at higher $T$, Eq~\ref{fit} can still fit the data because the Arrhenius factor is approaching its $E_{A}$-independent high-$T$ limit. Saddle dynamics, in which  $D(T)$ is \cite{inm}  proportional to the number of $Im-\omega$ instantaneous normal modes or \cite{jctk} the saddle order $<K(T)>$, apparently is consistent with $D \sim T$. Thus Eq~\ref{fit} can represent the full $T$-range, but the empirical activation energy is a compromise resulting from the fitting process with no simple relation to a barrier height.

In either case the crossover is seen to be a subtle, complex process. Application of the IS-mapping has revealed far more then could be deduced from $D(T)$ alone, including:
1. intrabasin and intrabasin {\it dominated} diffusion, associated with saddle dynamics
2. the distribution of local waiting times, well behaved as $N \rightarrow \infty$, calculated from the simulated distribution with the idea that the configuration space is a composite
3. a new, physically appealing criterion for dynamical crossover, $\tau_{loc} = \tau_{pl}$
4. a quantitative indicator $B(T)$ of the dominance of saddle or border dynamics (borderism), which also yields a crossover temperature.

The value of $T_{c}=0.52$, from a power-law fit to $D(T)$, is now just one of many pieces of evidence that a dynamical crossover occurs at $T \sim 0.50$. From recent prior work \cite{tkjc,jctk}: 1. IS energy $<U_{is}(T)>$ is undergoing its steepest fall. The true bottom is uncertain due to cooling-rate dependence and point 1. only identifies the general vicinity of the crossover. 2. saddle order $K(T)$ also extrapolates to zero at $T$ = 0.52 3. IST-Markov approximation becomes accurate at $T \sim T_{c}$. In this paper we have added: 4.  $D = D_{is}$, diffusion is intrabasin dominated, for $0.52 \geq T \geq 0.46$. 5. local waiting time reaches time to explore basin, $\tau_{loc} = \tau_{pl}$, at $T$ = 0.51. 6. border index $B(T)$ extrapolates to zero, $T$ = 0.53. We suggest that points 3 - 5 are particularly direct evidence that the mechanism of diffusion is changing.   They are based on ideas about the mechanism itself, while a fit to $D(T)$ produces a characteristic temperature based upon the 'symptoms'. Although 1 - 6 focus attention on a narrow range $0.53 \geq T \geq 0.46$ the crossover is gradual and might be thought to take place over $1.10 \geq T \geq 0.38$, the interval where intrabasin diffusion can be detected, encompassing the slope of $<U_{is}(T)>$. This agrees roughly with having the crossover begin at $T_{A}$.

We believe it significant that intrabasin diffusion can produce confinement for times long enough that the IST become a Markov chain, {\it without} invoking barriers. Thus diffusive confinement and confinement by barriers become two distinct features of the mechanism. Since intrabasin diffusion can be explained as saddle dynamics, our work fits with the growing recognition that the saddles of the landscape need to be considered on an even footing with the IS. There is no contradiction in having the intrabasin diffusion and IST-Markov regimes overlap. In the latter case $D \propto <\omega_{is}>$, and in the former, since escape from a basin requires diffusion to the border,  $<\omega_{is}> \propto D_{is}$.

Our simulation model exhibits most of the physical features currently being discussed for moderately supercooled dynamics, as well as some found here for the first time. There is much to be learned from LJ, $N$ = 32, although of course we intend to apply our methods to other liquids. It is difficult to achieve equilibrium below $T_{c}$. Note that, even though we report data down to $T$ = 0.34, only one significant $T$ appearing in the above estimates is below 0.50,  the end of intrabasin dominance at $T$ = 0.46. Our interest is the crossover and the lowest, probably non-equilibrated, $T$ are not required for the principal conslusions.

\section{Acknowledgment}
We would like to thank Frank Stillinger for valuable discussions. This work was supported by NSF Grants CHE9708055 and CHE0090975.


\begin{references}	
\par

\bibitem{angell0} C. A. Angell {\it Science} {\bf 267}, 1924 (1995).
\bibitem{dk} M. Ferrer, C. Lawrence, B. Demirjian, D. Kivelson, G. Tarjus and C. Alba-Simionesco, J. Chem. Phys. {\bf 108}, 8010 (1998).
\bibitem{ag} G. Adam and J. H. Gibbs, J. Chem. Phys. {\bf 43}, 139 (1965).
\bibitem{gold} M. Goldstein, J. Chem. Phys. {\bf 51}, 3728 (1969).
\bibitem{angell} C. A. Angell, B. Richards and V. Velikov, J. Phys. Condens. Matter {\bf 11} , A75 (1999).
\bibitem{goetze} W. Goetze and L. Sjogren, Rep. Prog. Phys. {\bf 55}, 241 (1992).
\bibitem{dk1} P. Viot, G. Tarjus and D. Kivelson, J. Chem. Phys. {\bf 112}, 10368 (2000).
\bibitem{sw} F. H. Stillinger and T. A. Weber, Phys. Rev. A {\bf 28}, 2408 (1983); {\it Science} {\bf 225}, 983 (1984).
\bibitem{fssci} F. H. Stillinger, {\it Science} {\bf 267}, 1935 (1995).
\bibitem{tkjc} T. Keyes and J. Chowdhary, Phys. Rev. E {\bf 64}, 032201 (2001).
\bibitem{sri} S. Sastry, P. Debenedetti and F. H. Stillinger, {\it Nature} {\bf 393}, 554 (1998).
\bibitem{andrea} A. Cavagna, Europhys. Lett. {\bf 53}, 490 (2001)
\bibitem{ang} L. Angelani, R. Di Leonardo, G. Ruocco, A. Scala and F. Sciortino, Phys. Rev. Lett. {\bf 85}, 5356 (2000).
\bibitem{bhup} B. Madan and T. Keyes, J. Chem. Phys. {\bf 98}, 3342 (1993).
\bibitem{lav} R. LaViolette and F. H. Stillinger, J. Chem. Phys. {\bf 83}, 4079 (1985).
\bibitem{tk} T. Keyes, J. Chem. Phys. {\bf 101}, 5081 (1994); Phys. Rev. E {\bf 59}, 3207 (1990).
\bibitem{jctk} J. Chowdhary and T. Keyes, Phys. Rev. E, submitted (2001).
\bibitem{jctk2} J. Chowdhary and T. Keyes, unpublished (2001).
\bibitem{jwalk} D. Frantz, D. Freeman and J. Doll, J. Chem. Phys. {\bf 93}, 2769 (1999); I. Andricioaei and J. E. Straub, J. Chem. Phys. {\bf 107}, 9117 (1997).
\bibitem{donati} C. Donati, F. Sciortino and P. Tartaglia, Phys. Rev. Lett. {\bf 85}, 1464 (2000).
\bibitem{inm} Wu-Xiong Li and T. Keyes, J. Chem. Phys. {\bf 111}, 5503 (1999); E. La Nave, A. Scala, F. Starr, F. Sciortino and H. E. Stanley, Phys. Rev. Lett, {\bf 84}, 4605 (2000).
\bibitem{kurchan} J. Kurchan and L. Laloux, J. Phys. A {\bf 29}, 1929 (1996).
\bibitem{andrea2} T. Grigera, A. Cavagna, I. Giardina and G. Parisi, preprint (2001).

\end{references}
\end{document}